# New biology of red rain extremophiles prove cometary panspermia

Godfrey Louis & A. Santhosh Kumar

School of Pure and Applied Physics, Mahatma Gandhi University,
Kottayam – 686 560, Kerala, India. (e-mail: `godfreylouis@vsnl.com`)

**This paper reports the extraordinary biology of the microorganisms from the mysterious red rain of Kerala, India. These chemosynthetic organisms grow optimally at an extreme high temperature of 300 degrees C in hydrothermal conditions and can metabolize inorganic and organic compounds including hydrocarbons. Stages found in their life cycle show reproduction by a special multiple fission process and the red cells found in the red rain are identified as the resting spores of these microbes. While these extreme hyperthermophiles contain proteins, our study shows the absence of DNA in these organisms, indicating a new primitive domain of life with alternate thermostable genetics. This new biology proves our earlier hypothesis that these microbes are of extraterrestrial origin and also supports our earlier argument that the mysterious red rain of Kerala is due to the cometary delivery of the red spores into the stratosphere above Kerala.**

**Keywords:** Astrobiology; extremophiles; panspermia; comet; red rain; hyperthermophiles; last common ancestor; origin of life; fourth domain of life.

## 1. Introduction

Living systems like microbes are well known for their ability to evolve and adapt to different environments, which are extreme from an anthropical point of view. Organisms that can live in extreme environments [1] have always been of great interest to astrobiologists. Finding of a number of hyperthermophilic microorganisms in hydrothermal vent systems and deep-sea oil wells [2-5] has shown that life can exist at temperatures near and above $100^o$ C. A bacteria isolated from the ocean vent was earlier reported [6] to be growing at $250^o$ C but this claim was debated later [7-9]. Until recently the accepted record for high temperature life was that of *Pyrolobus fumarii* [10] an archaea isolated from the ocean hydrothermal vent system that grows at $113^o$ C. This record was very recently broken by another claim [11] that report the finding of an ocean vent microbe called *Strain 121* that grow at $121^o$ C. Life on Earth as it is known to us today has serious limitations at temperatures much above this due to the physical and chemical limits set on the stability of biomolecules at high temperatures. These limits on higher temperature can be broken if microbes evolve to use alternate type of biomolecules or techniques to live in high temperature environment. At higher temperatures and pressures new type of chemical reactions and metabolic activities can be possible using alternate kind of biomolecules or techniques. Thus a theoretical upper temperature limit for life in the Universe cannot be defined by considering terrestrial life alone and there exists the possibility of finding life forms beyond Earth which can live in very extreme temperature and pressure conditions. Microbes can also evolve to live in very extreme chemical environments and can learn to extract energy for growth from such chemicals around them. Thus with the extension of its range, life can be a more universal phenomenon than presently thought.

We have shown in our earlier paper [12] that how the mysterious red rain phenomenon of Kerala, India, which occurred during July to September 2001, can be explained as due to stratospheric settling of the red cells from a fragile cometary body, which disintegrated in the upper atmosphere. While the red cells were found to be biological cells, their growth conditions were unknown. Stability of the red cells in extreme physical and chemical environments has prompted us to propose that the red cells are the spores of an extremophilic microorganism. In this paper we report the extraordinary growth conditions of these microorganisms. Experiments that prove the rapid and optimal growth of these microorganisms at an extreme high temperature of $300^o$ C are described. Present finding of life at $300^o$ C far exceeds all previously found and imagined upper temperature limits for life and opens a new era in biology.

All terrestrial living organisms are now known to contain DNA as informational macromolecules, which carry their genetic information. High temperature has deleterious effects on DNA [13]. The currently known hyperthermophiles are known to manage to live at temperatures near $100^o$ C using extensive DNA repair mechanisms against intense thermal denaturing [14]. But these types of mechanisms are doubtful to operate at extreme high temperature of $300^o$ C. Therefore an organism, which grows optimally at $300^o$ C, may use an alternate thermostable genetics. Here we are reporting our fluorescence study with ethidium bromide and chemical test using Diphenylamine to show that this organism is not containing DNA. This result is against the strongly established central dogma in terrestrial biology were DNA is the sole informational molecule that carry all genetic information. While this result may prompt any one to argue that these are not living systems we show that the absence of DNA is something, which can be expected from living organisms of this kind. We place these organisms in a newly proposed primitive domain of life called proto-domain.

In the light of the present results, we also present arguments to show that origin of life may not have taken place on Earth. We also discuss how life can spread in a galaxy from one generation of stars to another through cometary panspermia. A new mechanism is proposed for the launching of these organisms from planets to space, in which these organisms get multiplied in water containing planets around red giant stars and then finally get swept out to space by stellar outflow. Their presence in the interstellar space is speculated on the basis of their UV absorption feature, which correlates with the interstellar UV absorption peak at 217.5 nm. With these extraordinary evidences and arguments we claim that the red rain cells are of extraterrestrial origin. Thus we provide here a more clear answer to the age-old profound question on the existence of life beyond Earth.

## 2. Growth of red rain microbes

Studies conducted to investigate the growth conditions like temperature and substrate requirements shows that these organisms have a large temperature span for growth and they can also grow in a wide range of strange substrates. In combination with water the following materials like: transformer oil, cedar wood oil, petrol, glycine, sulphamic acid ($NH_2.SO_2.OH$) with $CO_2$, povidone iodine, ethanol, methanol, potassium permanganate and several other inorganic and organic compounds were found to support the growth of these organisms with varying degrees of activity. Silica gel, carbonates of calcium and magnesium were found to be growth aids. But sodium chloride is a deterrent for their growth. They can grow in total darkness and growth is not aided by light. In all substrates the growth rate of these organisms increases exponentially with temperature.

The growth characteristics of these microorganisms at very extreme conditions of temperature and pressure were investigated using a specially designed steel pressure vessel having a Teflon gasket. With this pressure vessel it is possible to go safely, without gasket blowout, up to a temperature slightly above $300^o$ C. A large number of experiments were conducted using this pressure vessel to understand the growth and reproduction techniques of these microorganisms. To find the optimum growth temperature the growth studies were performed at 100, 150, 200, 250, 300 and $350^o$ C using cedar wood oil and water as growth substrate. The temperature measurement was done using a PT100 Platinum RTD sensor. Growth was done in a 15 ml culture bottle kept in the pressure vessel, in which five drops of cedar wood oil and 10 ml of water was taken. Seeding was done with a wire loop. In order to minimize the growth at lower temperatures the pressure vessel was suddenly placed into the furnace, which was kept maintained at the required growth temperature. In all experiments the incubation time was 30 minutes. The sample was recovered from the pressure vessel after fast cooling. The growth was known both from the increase in turbidity and increase in number of cells visible under microscope. Turbidity measurement was taken using a spectrophotometer at 600 nm. The observation shows (Fig. 1) that the optimum growth is near 300° C. The possibility of a chemical artifact is ruled out because we have repeated the growth study at 300° C at least 50 times, often using totally unrelated chemicals as growth medium. Microscopic examination showed that same kind of cells were growing in all substrates.

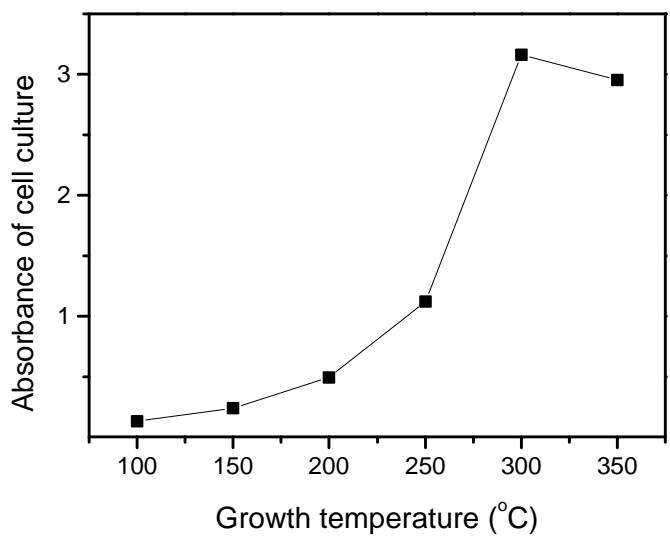

**Fig. 1.** Optimum growth temperature of red rain microbes. The curve shows the absorbance of the cell culture measured after the growth at different temperatures. Since the increase in absorbance was found to be due to increase in cell count, this curve shows that maximum growth is near 300° C. The upper temperature limit of these microbes are yet to be found.

To understand the nature of growth as a function of time the growth of these microbes were studied at fixed temperatures of 90 and 70° C. The growth was monitored using a computerized turbidity measurement set up. For this a white light LED source and a phototransistor detector were used along with a high-resolution (16-bit, Keithley Instruments) data acquisition card. The transmitted light intensity through the culture medium was measured every two minutes for several hours during the growth process. 1ml of 5% povidone iodine solution (Betadine) and 2ml of pure ethanol were added to 10ml of water to make the growth medium in a culture bottle. Using a heating arrangement the bottle was maintained at the required temperature. This growth medium was seeded with a wire loop using 300° C cultured sample. The growth curve (Fig. 2) shows typical microbe growth behavior having a lag phase, growth phase, a steady phase and a decline phase. The initial decrease in the light absorption curve is due to the color change of iodine from brown to colorless condition. This bleaching of brown color may be due to a chemical process. Exponential growth is noticed in the clear solution as an increase in turbidity after a lag phase. Also decline starts after few hundred minutes. Microscopic examination of the turbid culture showed that turbidity is due to the increased cell count. Control experiment was also carried out without seeding at 90° C and no growth was observed. Since the seeding was done using the 300° C grown sample, this experiment also demonstrated that the 300° C grown microbes are alive and are capable of further reproduction.

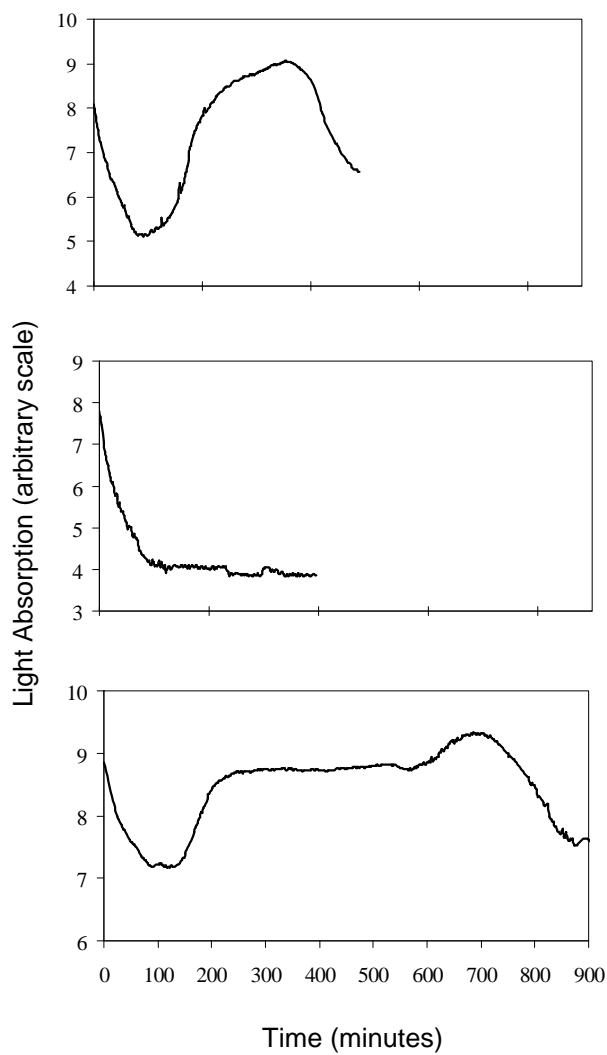

**Fig. 2.** Growth of microbes monitored at 90 and 70° C. Top panel shows the seeded growth at 90° C and middle panel shows the unseeded medium at 90° C where no growth is observed. Bottom panel shows the seeded growth at 70° C; the decline phase is delayed in this case due to the lower temperature.

Though these organisms grow optimally at 300° C, they can survive much above this temperature. The temperature at which the cells completely perish could not be found out due to the limitations of our pressure cell. In one case the Teflon gasket blowout occurred at $380^o$ C and the microorganisms could be grown from the traces left in the culture bottle. This showed that they do not perish even at $380^o$ C. At $373^o$ C water reaches a supercritical state with vapor having same density as liquid while the pressure reaches 220 atmospheres. Survival of biological cells in super critical water is unbelievable but these microorganisms some how manage to survive. The actual upper temperature limit of these microorganisms is yet to be found. There are some studies that show that biomolecular stability can be improved by high pressure [15,16]. Biology in supercritical water cannot be ruled out since the special structure of supercritical water has prompted Bassez [17] to suggest the possibility of origin of life in supercritical water.

**3. Life cycle**

The reproduction process of these microorganisms were initially unknown due to the difficulty of making direct observation at high temperatures and pressure, but by correlating a large number of observations of the cells after their growth at $300^o$ C, the following life cycles were identified. This finding was possible since some reproduction process was also found to take place at room temperature after their growth at $300^o$ C. The cells reproduce by a special process of multiple fission and three modes of life cycle have been observed for these microbes as shown in Figure 3.

*Cycle A* and *Cycle B* are reproduction processes and *Cycle C* is a spore forming cycle. Figure 4 shows photomicrographs of some of the different stages in the life cycle.

*Cycle A*: This is the fastest reproduction process of these microbes and is believed to be the major mechanism that operate at high temperatures. In this cycle some of the daughter cells of size 1micrometer grows to a size of about 4 to 6 micrometers (2nd stage of cycle A) and then finally into the size of about 30-50 micrometers (3rd stage of cycle A) and they attain a liquid bubble like structure. Several daughter cells are formed inside this liquid bubble and the bubble acts as a nursery of daughter cells (Fig. 4a). These daughter cells are finally released from the liquid bubble (4th stage of cycle A) leaving the used liquid bubble in the growth medium. This way the accumulation of these micro-liquid drops increases in the growth medium and it acts as a protective biofilm for the colony in extreme conditions.

*Cycle B*: This appears to be a slower process than cycle A and why some of the cells resort to this cycle is not clear. In this cycle some of the daughter cells grows to about 4 to 6 micrometers in size (stage 2 of cycle B) and then transform into a bubble like structure of mucus liquid as in cycle A to a size of 30 to 50 micrometers. This liquid structure then develops a shell of hardened material and its surface then becomes brownish yellow in color with a surface pattern somewhat similar to the dimple pattern of a golf ball ($3^{rd}$ stage of cycle B). This can be called a mother cell and several daughter cells develop inside this. When the daughter cells become sufficiently matured they start consuming the outer shell and a weak spot or a protrusion develops in the mother cell. Through this protrusion daughter cells are released from the mother cell leaving the dimple patterned empty globes in the growth medium (stage 4 of cycle B). In one rare observation about 15 to 20 daughter cells were found to escape from a mother cell. The daughter cells start their life as tiny cells of about 1 micrometers in size and grow to larger size to repeat the life cycle.

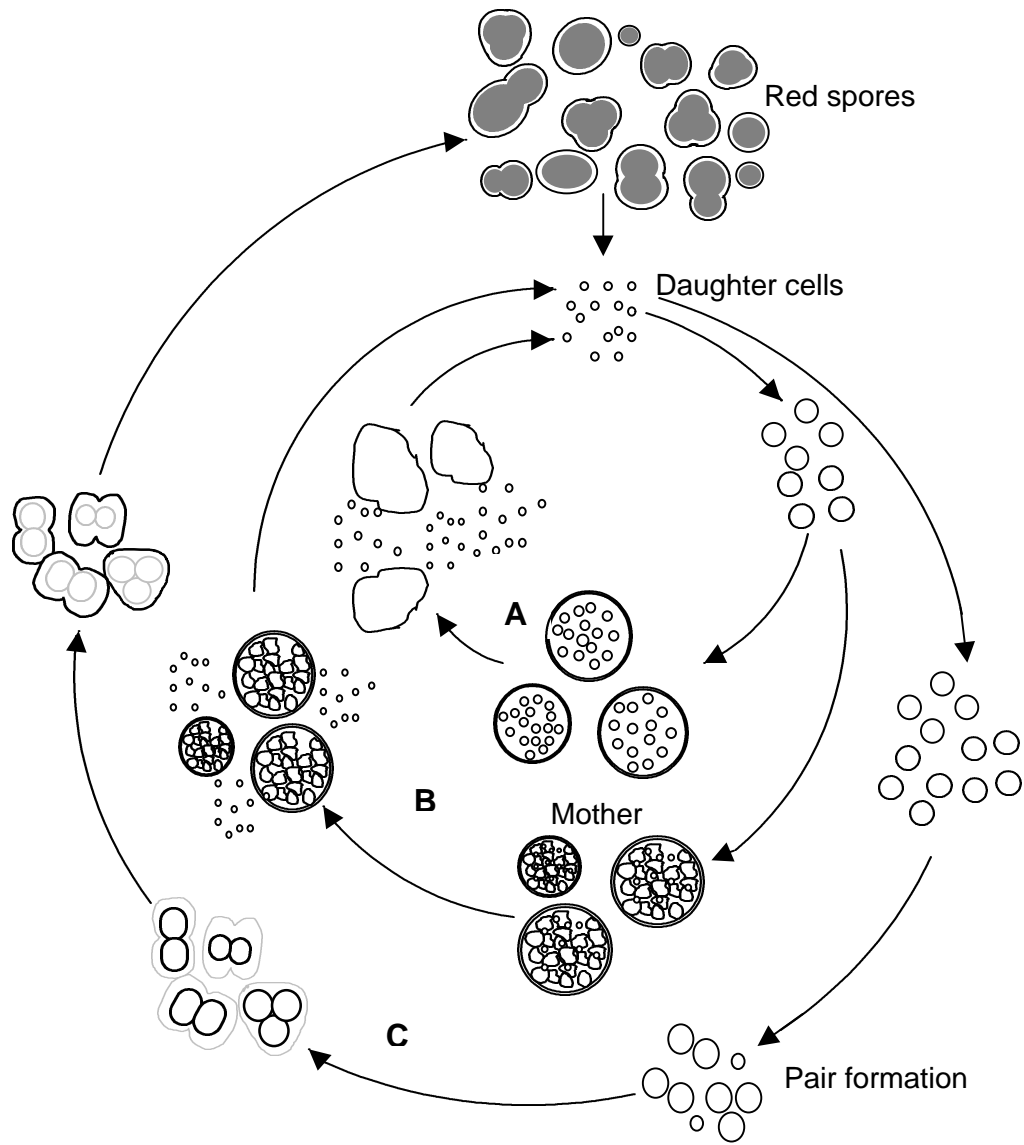

**Fig. 3.** Life cycles of red rain microbes. Figure shows two kind of reproductive and one kind of spore forming cycle of the red rain microbes. Cycle A is a fast reproductive process in which several daughter cells are generated by multiple fission inside a large sphere of mother liquid. In cycle B the mother liquid develops a cell wall and the daughter cells emerge through a protrusion in the wall of the mother cell. Cycle C is the spore forming cycle and it is a slow process. (The cell sizes are not drawn to scale).

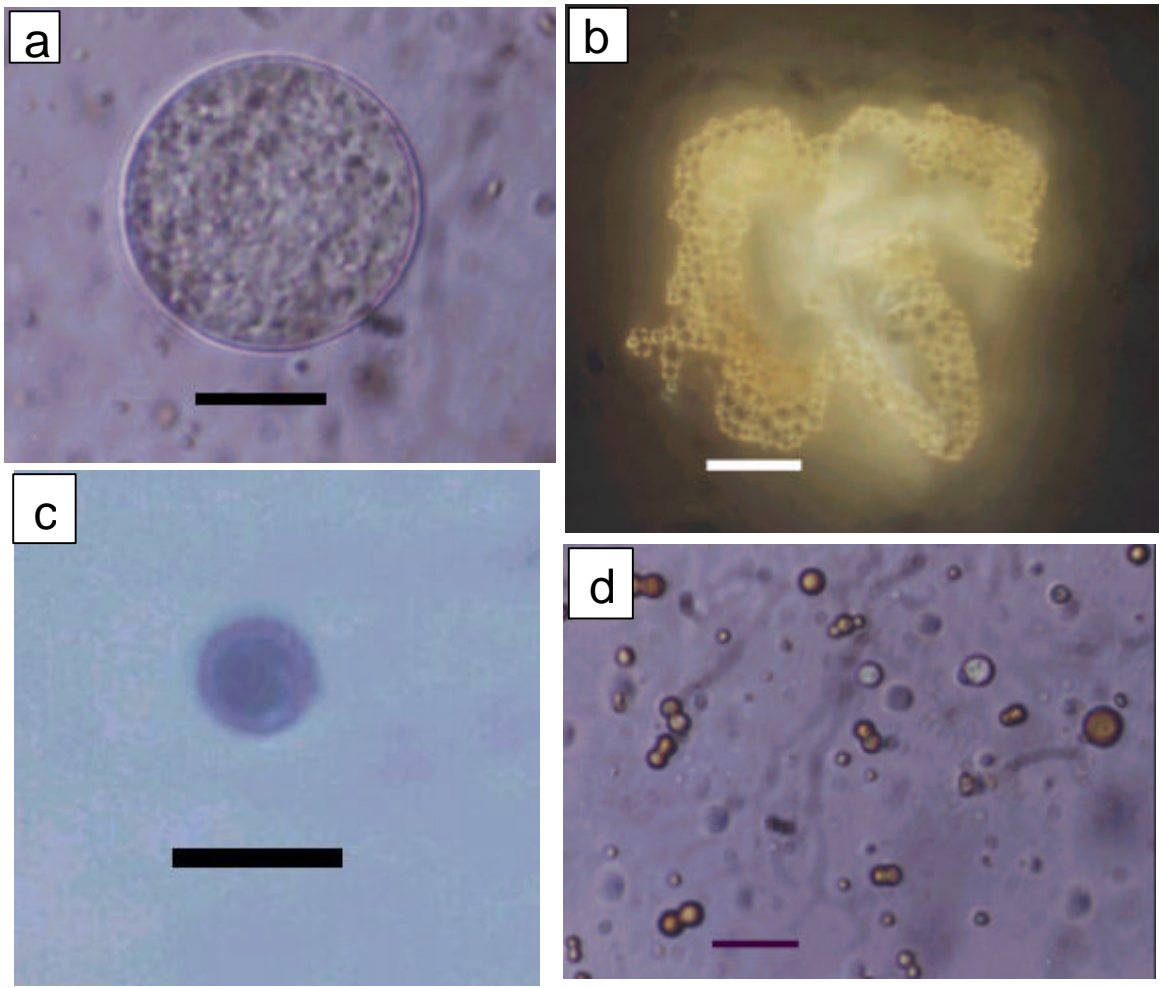

**Fig. 4.** Photomicrographs of grown cells. A) A bubble like structure of mother liquid which contains several small daughter cells. These daughter cells can be seen to perform an intense struggle inside the mother liquid when they get matured enough to get out. B) A dried cluster of daughter cells photographed under reflected light through crossed polarizers. This indicates optical activity in these cells caused by the biomolecules. C) A single grown cell which has absorbed dye. The layered structure of the cell is visible after dye penetration D) This image shows the cells, which are proceeding to form spores. Many cells forms pairs or triplets before entering the spore state.  Scale bars indicate 10 $\mu$m.

*Cycle C*: This is the spore forming life cycle of these microbes. Prolonged growth at high temperature with lack of nutrients appears to be the reason for spore formation. In this cycle, some of the grown cells can be found to show a pair formation tendency (stage 3 of cycle C) as a first step towards the formation of spores (Fig. 4d). Two or three or rarely even more cells linger around each other closely for some time and they finally get fused together and form a common thin outer mucus layer around them (stage 4 of cycle C). Later this thin mucus layer becomes very thick and hard (stage 5 of cycle C) and the compound cell becomes a spore with color change from white to yellowish and finally red (stage 6 of cycle C).  The red spore cells show different shapes like spherical, ellipsoidal and slightly elongated shapes with septum like formation at the center and also some have a triangular shape.  The elongated shapes are due to the fusion of two cells of equal or unequal sizes and the triangular shapes are due to the fusion of three cells. The original red rain cells that caused the red rain phenomenon in Kerala also exhibit these characteristic shapes. This spore state appears to be a resting phase of these organisms.  The thick outer layer of the spores disintegrates and releases the enclosed original cells only when nutrients and growth conditions are persistently available. These spores can possibly remain dormant in comets and interstellar clouds for millions of years till they reach suitable environments like hot planetary bodies around newly born stars. Their thick outer layer is possibly a sufficient protection against UV and other radiation in space.

## 4. Test for biomolecules

The absence of DNA in the cells was found both by chemical and spectroflourometeric techniques. For chemical testing the method of Dische Diphenyalamine reagent was used. 2 ml of the Diphenyalamine reagent was added to 1 ml of the cultured cell suspension and the mixture was heated in a boiling water bath for 10 minutes. The characteristic blue color that indicates the presence of DNA could not be observed.  However a deep green color was noticed which cannot be attributed to DNA.  Similar tests were performed for yeast cells and blue color was clearly observed indicating the presence of DNA. Again the absence of DNA in red rain microbes were confirmed by spectroflourometeric technique using ethidium bromide fluorescent dye. The fluorescence emission of ethidium bromide at 600 nm is enhanced greatly in the presence of DNA. The test for the DNA is performed using this property. For this the cells were first centrifuged out from the suspension and were well crushed in a mortar. Two drops of this cell extract was added to ethidium bromide stock solution taken in a cuvette.  Fluorescence emission spectrum of this mixture was recorded using a Shimadzu spectrofluorophotometer (model no. RF – 5301 PC).  This spectrum did not show an enhanced fluorescence as expected from a DNA containing sample.  The experiment was repeated after grinding the cell in liquid nitrogen to ensure the cracking of cells. This also showed no enhancement in fluorescence indicating the absence of DNA in these cells. These tests were also performed on the original red rain cells and in this case also no DNA could be detected. Similar experiments when performed in yeast cells and other plant materials showed greatly enhanced fluorescence of the ethidium bromide indicating the presence of DNA molecules in these samples.

 The presence of protein was tested using Xanthoproteic test. For this the cultured cells were heated with conc. Nitric acid. On adding sodium hydroxide to this mixture a deep yellowish orange color developed, indicating the presence of proteins. The cells are capable of absorbing various dyes like Methyl Green, Crystal Violet and Ethidium Bromide into the cell body and get stained. This is an indication for the presence of dye binding biomolecules. The layered structure of the cells appears more clearly under staining (Fig. 4c).

## 5. UV absorption study

In UV visible spectrophotometer study (using Shimadzu model no. UV – 2401 PC), even very dilute suspensions of these cells in water showed a strong UV absorbing characteristics. The absorption spectra of the microbes cultured at 200 and $300^o$ C are shown in Figure 5. These curves show an absorption peak near 200 nm. An increase in absorbance for growth at a higher temperature is due to the increased number density of cells. The above UV absorbance characteristic is exhibited by all the suspensions of these organisms irrespective of the widely different substrates in which they are grown. This spectral characteristic is very similar to the absorption spectra of original red rain sample [12], which are actually the spores of these microorganisms. This strong UV absorption characteristic gives these organisms a possible protection in space environment from harmful UV radiation. An outer layer of sacrificing cells can give efficient protection to the inner layer of cells in a cometary body.

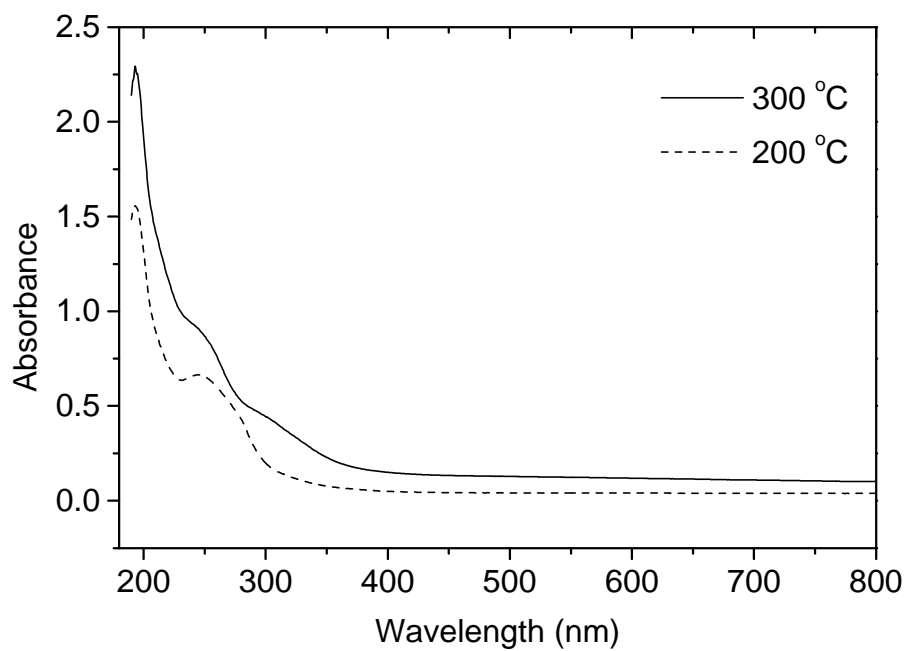

**Fig. 5.** The UV-Visible absorbance spectrum of the cultured red rain microbes in water. To record this spectrum in the linear range of the instrument (which goes only up to 5 in the logarithmic absorbance scale) the original culture was diluted by a factor of 32. This indicates the very strong nature of the UV feature of these microbes. The solid and doted lines indicate two different culturing temperatures as indicated.

## 6. Discussion

According to the study presented in our earlier paper [12] the red rain phenomenon was explained as due to cometary panspermia. The present findings strongly prove this idea. The microorganism isolated from the red rain of Kerala shows very extraordinary characteristics like ability to grow optimally at 300° C and the capacity to metabolize a wide range of organic and inorganic materials. The absence of DNA in the red rain microorganisms and their ability to grow optimally at this extreme high temperature of 300° C and their capacity to survive temperatures near 400° C are a definite indication for their extraterrestrial origin. Because these characteristics are not exhibited by any known terrestrial organisms.

*Case of DNA:* It can be argued that our test for DNA has failed because of some unusual binding of DNA with some molecules in these organisms that interfere with these tests. But the following observations show that this cannot be the case. One of the main chemical constituents of the DNA molecule is phosphorus. In our growth experiments intense growth can be observed without the aid of any phosphorus compounds and no growth enhancement could also be observed on adding phosphorus compounds. This is another reason for us to argue that this organism is not having DNA. The crucial 6 elements for terrestrial like life are H, C, N, O, S & P. Of these elements P is the only element, which has not yet been identified in comets [18]. This also points to the possibility that comet dwelling organisms may not have a P containing genetic molecule. Our growth studies indicate that the cells are mainly built with elements C, H, O, & N and the compounds containing these elements are abundant in comets and interstellar clouds [18,19]. Further, as a survival strategy to exist in space these organisms may have evolved without DNA to prevent accumulated radiation damage [20] due to low radiation from radioactive materials in a comet or from cosmic rays. Further, it is also quiet possible that organisms that have to adapt to very high temperatures like 300 to 400° C may avoid thermally unstable DNA as their genetic molecule. There are some speculations [21] that consider prebiotic polyamino acids as the first informational macromolecules. Since proteins are more thermostable [22, 23] than DNA or RNA there exists the possibility that a smart collection of special type of thermostable informational proteins with their complex set of interactions may be controlling the cellular functions of these extreme hyperthermophilic organisms. Recently Nelson, Levy and Miller [24] have suggested that peptide nucleic acids (PNA) rather than RNA may have been the first genetic molecule. The genetics based on PNA or a similar molecule is a possibility in the present case because PNA do not have a phosphate-based backbone.

***Microbes in interstellar clouds:*** The following discussion based on the UV absorption characteristics of these microbes show why their presence in interstellar medium can be suspected. One of the most relevant signatures of interstellar dust is the 'extinction' of starlight, caused by this dust, towards different sources in the Galaxy. The reason for the UV extinction bump at 217.5 nm [19, 25] is not well understood. Hoyle, Wickramasinghe and others [26-28] have earlier considered the presence of biological grains in space and its possible relation to the interstellar UV extinction bump at 217.5 nm. But one objection [29, 30] to this idea was that there is no cosmic abundance of the element P to assume the presence of microorganisms in space. Our finding overrules this objection because red rain microbes have no DNA and hence no P requirement for their growth. The red rain microbes, both their spores and vegetative cells, show a strong peak in UV absorption near 200 nm. As speculated in our earlier paper [12], the UV extinction bump at 217.5 nm of the interstellar dust can be correlated with this by assuming that the red rain microbial spores are present as a component in the interstellar dust clouds. We have measured the UV absorbance with the cells suspended in water at room temperature. But in space the cells are suspended in vacuum at low temperature. Thus a shift of about 20 nm in the absorption peak can possibly be attributed to the effect of low temperature and pressure condition in the interstellar space on the dehydrated spores of the red rain microbes. Further studies of the red cells in a simulated interstellar condition can test this speculation.

***Origin of life:*** Did life had an independent origin on Earth? If not, did it come here from space? These questions can be examined in the light of the present findings. Biosignatures [31–33] and microfossil [34, 35] records indicate that life made a very early appearance on Earth. The currently favored theory by a majority of scientists is that this life originated on Earth from precursor organic molecules. While the formation of biomolecules using cometary [36, 37] or terrestrially formed organic compounds [38] can be a real possibility, its assembly into a living cell with a complex genetic program through random trials appears astronomically improbable in a relatively short geological time period. The question of how the specific information content in biological cells originated on Earth in a short time is a more difficult question than the question of how the biomolecules like proteins and DNA first originated here. Those who propose that life is a result of intelligent design [39] strongly raise this question of the origin of information to defeat the theory of chemical origin of life. The present finding gives a very natural explanation for the rapid appearance of life

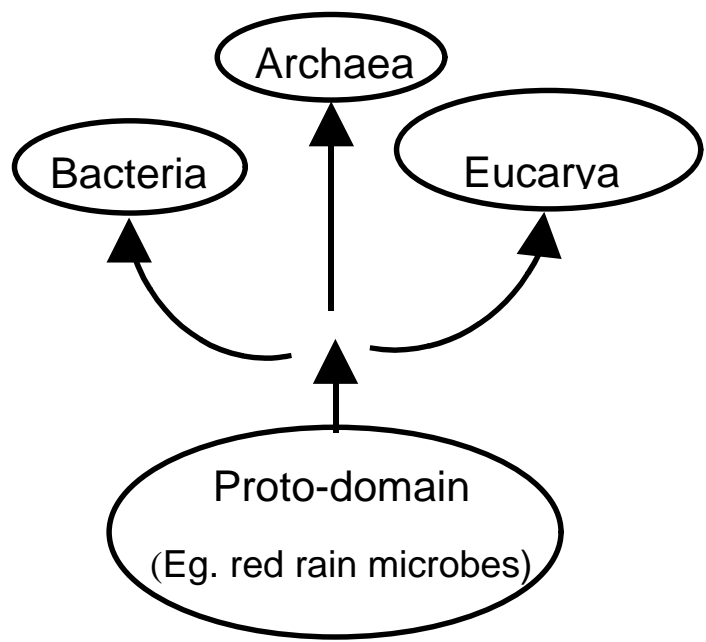

**Fig. 6.** A new domain of life. The extreme hyperthermophilic nature and the absence of DNA in the red rain microbes justifies their placement in a new domain, which may be called proto-domain, from which all the currently identified three domains of life may have emerged on Earth. Cellular functions of proto-domain organisms are possibly controlled by a smart collection of proteins. DNA and RNA are considered as products of evolution in this scheme of life. Absence of DNA in red rain microbes need not be an indication that they are lesser beings than bacteria because avoiding DNA can be an evolutionary necessity for their space travel capability

on Earth. Several studies indicate that the last common ancestor of all extant life on Earth is a hyperthermophilic organism [40-42]. This first DNA based ancient thermophilic organism or organisms from which the tree of life diverged may have evolved from the extreme hyperthermophilic red rain organisms in the early stages of Earth. Thus the DNA based life, which is presently observed on Earth can only be a case of lower temperature adapted biological evolution possible from the red rain organisms. Thus DNA can be a product of evolution rather than a prebiotic molecule [43, 44]. These arguments lead to the conclusion that red rain organisms may represent a primitive domain of life, which may be named as proto-domain, from which the presently known three domains of life [45] may have emerged (Fig. 6.). It can be observed that the organisms placed towards the root of the tree of life are increasingly extremophilic and if this is taken as a rule, then placing the red rain organisms below the root of the tree of life is most appropriate. Placing them in a newly proposed domain is also justified because they are not DNA based organisms.

Now consider the early earth period of nearly 4 billion years ago. The Earth was under heavy cometary bombardment during its early period [46]. Only extreme hyperthermophilic and chemosynthetic organisms like red rain microorganism can survive comfortably at temperatures as high as 300 to 400° C in an impact heated steaming and cloudy early earth environment. The life on Earth must have started with the seeding through the red rain process caused by some of the impacting comets. These arguments show that origin of life may not have taken place on Earth.

*Life distribution mechanism:* Considering the fact that the red rain organisms can undergo large scale multiplication at high temperatures gives the possibility for a new mechanism for their wide distribution in a galaxy. It can be seen that a large-scale amplification of these organisms can take place in water containing planets orbiting the red giant stars. A recent simulation study [47] shows that planets with large water content can be very common around stars. During the final stages in the lifetime of a star such as the red giant stage, the size and luminosity of a star increases and the temperature in orbiting planets increases to high levels. This will lead to the extinction of organisms or creatures present in the planet, which are adapted to cooler temperatures. A water-containing planet can now become a hot steaming place, where an enormous multiplication of the dormant red rain organisms can take place. Their growth will continue till all the possible nutrients on a planet have been utilized and finally the production of red spores will take place under nutrient depletion. These spores can remain suspended in the hot atmosphere of dense water vapor created by the evaporation of all the oceans in the planet. As stars like sun approaches the red giant stage a large fraction of their mass is thrown away to interstellar medium as thick clouds of gas and dust through high-speed stellar winds [48]. These high-speed dense stellar winds can sweep away the spores and water vapor from the planetary surface. This process is aided by the fact that planets loose their protective magnetosphere due to core freezing as they get old. Considering the high water holding capacity of planets, the amount of spores escaping from the planets can be astronomically large. Finally the red spores can get distributed in the interstellar space by being propelled by radiation pressure. These red spores can then get incorporated later into new comets and then seed new planets, which are formed around the next generation of stars. Thus this mechanism can allow the continuation and distribution of life throughout a galaxy from one generation of stars to another. This also explains why these cells can be expected as a component of interstellar dust and consequently it can give rise to the UV extinction feature at 217.5 nm.

The speculations discussed above which are backed by the present evidence shows that life is not constrained to Earth alone and it can be present in interstellar dust clouds and other planetary and cometary bodies. Thus many unexplained or poorly explained astronomical observations are to be reinterpreted in terms of possible biological cause.

Present finding also gives the possibility for the active existence of these organisms in a nearby hot place in the Solar System like planet Venus.

## Acknowledgements

G.L. is the main author of this work. A.S.K. has made contribution by providing great assistance to G.L. for the conduct of experiments connected to this work. We greatly acknowledge the help of George Varughese for collecting many of the red rain samples for study. We thank Dr. Sabu Thomas for providing photomicrography facility. We also thank the teaching, administrative and technical staff members, research scholars and students of School of Pure & Applied Physics and several others who have encouraged and helped this work.